\documentclass[twocolumn,showpacs,amsmath,amssymb,prl]{revtex4}
\usepackage{graphicx}
\usepackage{dcolumn}
\usepackage{bm}
\begin{document}

\title{Superconductivity and topological Fermi surface transitions in
electron-doped cuprates near optimal doping}
\author{Tanmoy Das, R. S. Markiewicz, and A. Bansil}
\address{Physics Department, Northeastern University, Boston MA
02115, USA}
\date{\today}
\begin{abstract}
We discuss evolution of the Fermi surface (FS) topology with doping
in electron doped cuprates within the framework of a one-band
Hubbard Hamiltonian, where antiferromagnetism and superconductivity
are assumed to coexist in a uniform phase. In the lightly doped
insulator, the FS consists of electron pockets around the $(\pi,0)$
points. The first change in the FS topology occurs in the optimally
doped region when an additional hole pocket appears at the nodal
point. The second change in topology takes place in the overdoped
regime ($\sim18\%$) where antiferromagnetism disappears and a large
$(\pi,\pi)$-centered metallic FS is formed. Evidence for these two
topological transitions is found in recent Hall effect and
penetration depth experiments on Pr$_{2-x}$Ce$_{x}$CuO$_{4-\delta}$
(PCCO) and with a number of spectroscopic measurements on
Nd$_{2-x}$Ce$_{x}$CuO$_{4-\delta}$ (NCCO).
\end{abstract}
 \maketitle \narrowtext

\section{Introduction}
An understanding of the origin of electron-hole asymmetry can
provide important clues for unraveling the mechanism of pair
formation in high$-T_c$ superconductors. Our recent
analysis\cite{kusko,tanmoy,tanmoy2} indicates that the electron
doped cuprates behave like uniform antiferromagnetic metals and
superconductors over a wide doping range up to a quantum critical
point (QCP). As the half-filled state is doped, electrons condense
into pockets around the $(\pi,0)-$points at the bottom of the upper
magnetic band (UMB). With increasing doping the magnetic gap
decreases, and as the lower magnetic band (LMB) crosses the Fermi
level ($E_F$), hole pockets appear near the nodal regions, resulting
in the first topological transition (TTI) of the Fermi Surface (FS).
Here the nodal hole pockets coexist with the $(\pi,0)$ electron
pockets, separated by the hot-spot regions of the FS due to residual
antiferromagnetism. With further electron doping, antiferomagnetic
(AFM) order is destroyed as the magnetic gap collapses around
$x\approx 0.18$, and the FS crosses over from being a collection of
small pockets to a large metallic $(\pi,\pi)$-centered sheet,
yielding the second topological transition (TTII) of the
FS\cite{foot6}. Our model of these two topological transitions in
the FS is consistent with angle resolved photoemission spectroscopy
(APRES)\cite{armitage}, resonant inelastic X-ray scattering
(RIXS)\cite{markie} and other experimental results as discussed
below.

It is interesting to ask how the aforementioned topological
transitions manifest themselves in the superconducting properties of
the cuprates. In this connection, we show that the FS topology is
reflected directly in the $T-$dependence of the penetration depth
$\lambda$: the behavior of $\lambda$ crosses over from an apparent
$'s'-$wave (nodeless $d-$wave) form below TTI, to a mixed
$d+'s'-$wave form above TTI, to a pure $d-$wave form above TTII,
even though the underlying pairing symmetry remains $d-$wave at all
dopings. The present article thus expands on our earlier penetration
depth study\cite{tanmoy2}, which showed that a linear-in-$T$
superfluid density $n_s\propto\lambda^{-2}$ originates in the hole
pockets around the nodal points, and that the superfluid density of
the electron pockets varies exponentially with $T$. $n_s$ then
appears nodeless in the underdoped regime due to strong AFM
correlations. Near optimal doping, the appearance of the nodal
pocket produces gapless hole quasiparticles which dominate at low
$T$. Interestingly, with increasing $T$ the small gap on the hole
pocket is destroyed by thermal excitations and the system is left
with a gap only on the electron pockets. The interplay between AFM
order and superconductivity (SC) in determining how the SC pairing
symmetry manifests itself in our calculations resolves a long
standing controversy regarding the pairing symmetry in electron
doped
cuprates\cite{anlage,kashiwaya,pcco,chesca,snezhko,ariando,skinta,
biswas,qazilbash}.

Recent Hall effect measurements\cite{greene1,greene2} provide
further evidence for the existence of the TTI involving the
appearance of nodal hole pockets around optimal doping. We show in
this article that, above optimal doping, the experimentally observed
crossover from a positive Hall coefficient at low $T$ to negative at
higher $T$ is a direct consequence of the coexistence of two types
of charge carriers. We also comment on possible consequences of the
coexistence of the electron- and hole-like quasiparticles such as
non-Fermi-liquid behavior\cite{krotkov} and Bose-Einstein
condensation\cite{brinkman,adhikari}.
\section{Mott Insulator to Mott gap collapse}
\begin{figure}[tp]
\rotatebox{270}{\scalebox{0.36}{\includegraphics{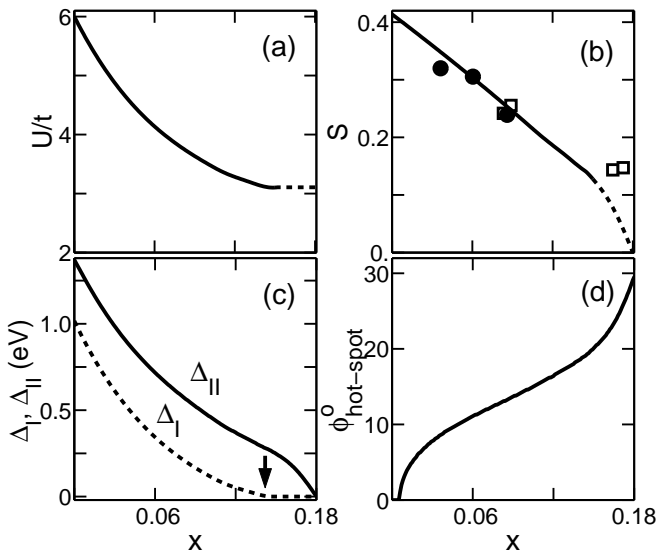}}}
\caption{ (a) Doping dependence of $U/t$. (b) Self-consistently
computed magnetisation, $S$, (solid line, extrapolated and shown
dashed at higher dopings as it terminates in a QCP) and the
corresponding experimental results from
Ref.~\protect\onlinecite{mgreven} (open squares) and
Ref.~\protect\onlinecite{rosseinsky} (filled dots). (c) Nodal point
pseudogap ($\Delta_{\rm{I}}$), defined as the energy gap from the
$E_F$ to the LMB at the nodal point, is plotted as a function of
doping in dashed line. The arrow marks the doping ($x=0.145$) where
this pseudogap vanishes at the first topological transition TTI.
Hot-spot gap ($\Delta_{\rm{II}}$) is shown in solid line which
terminates near 18\% doping, denoting the second topological
transition, TTII. (d) The position of the hot-spot gap is given by
the FS angle, $\phi_{\rm{hot-spot}}$; the angle is zero (45$^o$)
along the antinodal (nodal) direction.}\label{param}
\end{figure}
\begin{figure}[tp]
\rotatebox{270}{\scalebox{0.40}{\includegraphics{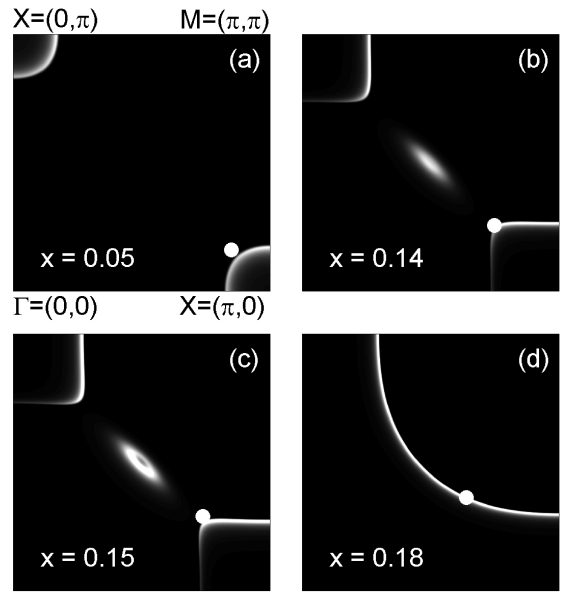}}}
\caption{Computed spectral intensity at the Fermi energy in PCCO,
which gives an impression of the FS at different dopings. In (b),
even though the LMB is below the $E_F$, some intensity can be seen
around the nodal point due to the integration of the spectral
intensity over a finite energy window around $E_F$ (to mimic
experiment). Whites denote high and blacks low intensity. The dots
mark the positions of the hot-spot as discussed in
Fig.~\protect\ref{param}(c)-(d); and the dot in (d) represents the
limiting value of momentum where hot-spot disappears} \label{fermi}
\end{figure}
Our analysis proceeds within the framework of a one-band Hubbard
Hamiltonian with tight-binding (TB) hopping parameters $t,
t^{\prime}, t^{\prime\prime}$, on-site repulsive interaction $U$,
and a $d-$wave pairing interaction $\Delta$; see
Refs.~\protect\onlinecite{tanmoy} and \onlinecite{tanmoy2} for
details. Mean field Hartree-Fock approach is used to solve for all
the order parameters self-consistently in the AFM, normal as well as
the SC states, assuming that the effective $U$ is doping dependent
as shown in Fig. 1(a). $U$ varies rapidly at low dopings, but the
doping dependence is much weaker at higher dopings\cite{foot6}. A
finite expectation value of the staggered magnetization $S$ at the
nesting wave vector $\vec{Q}=(\pi,\pi)$ produces a gap in the low
energy spectrum in a mean field treatment. $S$ is seen from Fig.
1(b) to decrease linearly with doping up to optimal doping and to
then drop sharply, disappearing around $x=0.18$, even though $U$ is
nearly doping independent at higher dopings. Incidentally, we adduce
that the values of the TB parameters and $U$, and thus of $S$ are
quite similar in NCCO\cite{kusko} and PCCO\cite{tanmoy2}, although
Pr$_{1-x}$LaCe$_{x}$CuO$_{4-\delta}$ (PLCCO) seems to be somewhat
different\cite{tanmoy}. The magnetic gap given by $US$ splits the
bare band into the upper (UMB) and lower (LMB) magnetic bands, where
at half-filling, the LMB is fully occupied and the UMB is empty. A
gap between the $E_F$ and the top of the LMB at the nodal point
would result in a nodal pseudogap ($\Delta_{\rm{I}}$) in the
spectrum when fluctuation effects missing in our mean field
computations are accounted for. This nodal pseudogap is seen from
Fig. 1(c) (dashed line) to have a large value of 1.0 eV at half
filling, and to close at the TTI. In contrast, the pseudogap at the
hot-spot ($\Delta_{\rm{II}}$), computed by the maximum gap in the
FS, is larger than the nodal pseudogap (shown by the solid line),
and closes at the QCP (i.e. TTII). The position of the maximum
hot-spot gap is represented by the FS angle
($\phi^o_{\rm{hot-spot}}$) in Fig. 1(d), where the angle is zero
(45$^o$) along the antinodal (nodal) direction. Interestingly, the
hot-spot in electron doped cuprates appears in the antinodal
direction at half filling. A similar behavior is also seen in the
hole doped case\cite{tanmoy2gap}. With increasing doping, the gap
moves away towards the nodal direction nonlinearly with doping,
where for hole doped cuprates, it stays at the antinodal direction
for all dopings. The white dots in Fig. 2 mark the same
$\phi^o_{\rm{hot-spot}}$ on the FS maps at different dopings.

Turning to the computational details of the doping evolution of the
FS, when the insulator is doped with electrons, the FS first forms
as small, nearly circular electron pockets centered at
$X=(\pi,0)/(0,\pi)$, leading to an AFM metal as shown for $x=0.05$
in Fig. 2(a). With increasing doping, these electron pockets become
more squarish in shape, and lose spectral weight along the $\Gamma
\rightarrow X$ direction as seen in Fig. 2(b). As we approach
optimal doping, the nodal pseudogap vanishes around $x=0.145$
(marked by the arrow in Fig.\ref{param}(c)) as the LMB crosses
$E_F$, producing the necklace-like FS of Fig. 2(c). Finally, a
complete metal-like FS is restored in Fig. 2(d) as the magnetic gap
collapses around $x=0.18$.
\section{Superconductivity and Penetration Depth}
\begin{figure}[tp]
\rotatebox{270}{\scalebox{0.40}{\includegraphics{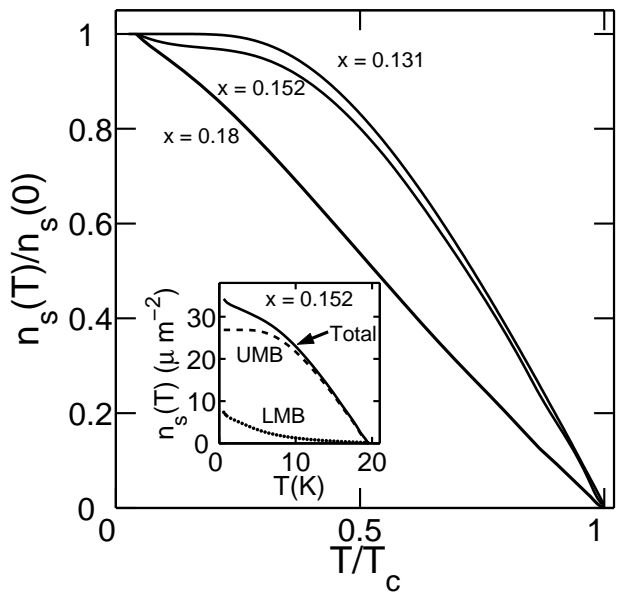}}}
\caption{Computed values of the normalized superfluid density
$n_s(T)/n_s(0)$ as a function of scaled temperature $T/T_c$ at three
different dopings in PCCO \cite{pcco}. $n_s(0)$ is the superfluid
density at $T=0$ and $T_c$ is the SC transition temperature. {\it
Inset}: Total $n_s$ at $x=0.152$ is shown decomposed into UMB and
LMB contributions.} \label{penet}
\end{figure}
\begin{figure}[tp]
\rotatebox{270}{\scalebox{0.40}{\includegraphics{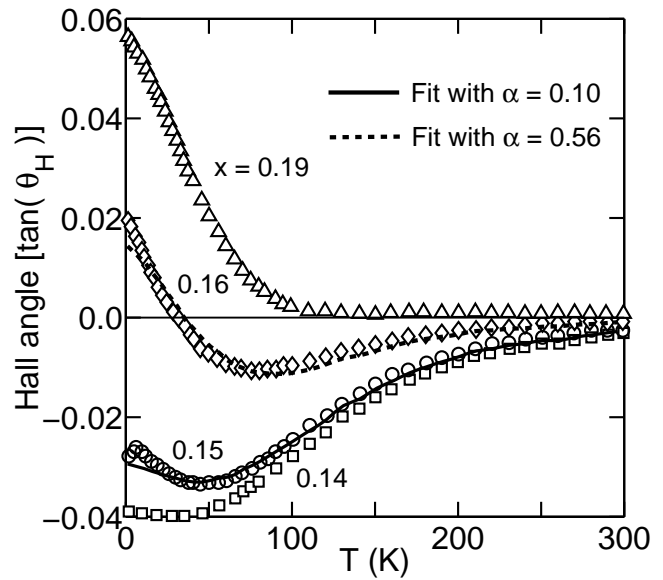}}}
\caption{ Experimental values of the Hall angle $\theta_H$
\cite{greene2} as a function of $T$ at several dopings. Solid and
dashed lines through the data for $x=0.15$ and $x=0.16$,
respectively are qualitative fits of form [$\alpha \theta_H (x=0.19)
+ (1-\alpha)\theta_H(x=0.14)$], where $\alpha$ is a mixing
parameter.} \label{hallangle}
\end{figure}
The superfluid density $n_s\propto\lambda^{-2}$ is a fundamental
property of the superconductor and its evolution with doping
provides insight into the interplay between the AFM and SC orders.
The low$-T$ behavior of $n_s$ is traditionally used to determine the
pairing symmetry\cite{prohammer}, but results in electron-doped
cuprates have been contradictory. Some early measurements found
evidence for $s-$wave pairing\cite{anlage,kashiwaya}, but other
tunneling\cite{chesca} and penetration depth\cite{snezhko,ariando}
experiments report $d-$wave pairing. Yet other experiments suggest a
transition from an $s-$wave in underdoped samples to either a
$d-$wave\cite{skinta,biswas} and/or a mixed
$(d+is)-$wave\cite{qazilbash} character in the optimally and
overdoped cases.

We have addressed this issue by directly calculating the penetration
depth in a model with coexisting AFM and SC orders, assuming {\it
$d$-wave pairing at all dopings}. The technical details of the
calculations and the corresponding doping dependent SC order
parameters are given in Ref.~\onlinecite{tanmoy2}. Typical results
for the superfluid density $n_s$ are given in Fig. 3, and show that
it varies exponentially in the underdoped region at $x=0.131$, even
though the pairing symmetry is $d-$wave. This is due to the absence
of spectral weight in the nodal region associated with the large
nodal pseudogap. In contrast, in the optimal doping region at
$x=0.152$\cite{foot5}, $n_s$ shows a linear-in-$T$ behavior in the
very low$-T$ region ($T < 1.5K$) as the nodal hole pocket is formed
(see inset in Fig. 3). Interestingly, for higher $T$, the
theoretical penetration depth shows a transition from a
linear-in-$T$ to an exponential in accord with experimental
observations\cite{pcco}.

To clarify the complicated $T-$dependence of $n_s$, we have
investigated the partial contributions to the total superfluid
density from the UMB and LMB at $x$=0.152 and the results are
summarized in the inset in Fig. 3. Since the UMB lies well above the
nodal point at all dopings, the associated contribution displays an
apparent $'s'-$wave like behavior (dashed line in the inset). Once
the nodal pocket is formed the LMB contributes a linear-in-$T$
behavior (short dashed line). But at $x$=0.152, this hole pocket is
quite small, and hence the linear-in-$T$ behavior persists only at
very low $T$. At high $T$ the electron quasiparticles dominate,
yielding to an $'s'-$wave-like plateau in the total superfluid
density (solid line). This leads to an apparent mixed $d+'s'-$wave
like behavior in the pairing symmetry.

A second transition is expected in the overdoped regime near
$x=0.18$, where the magnetic gap collapses and only a single large
FS sheet remains. This transition should be clearly observable in
penetration depth as seen from the dashed curve in Fig. 3 in which
there is no trace of the $'s'-$wave like plateau. So far, however,
experimental data has to our knowledge not been reported on such
highly doped samples. It is clear thus that the doping evolution of
$n_s$ provides a direct handle on the topology of the underlying FS.
\section{Transport Properties}
Dagan {\it et. al.} \cite{greene1,greene2} have shown in PCCO that
the Hall coefficient (see Fig.~ 4), $R_H$, as well as the Hall
angle, $\theta_H$, defined as $\tan( \theta_H
)=\rho_{xy}/\rho_{xx}$, where $\rho_{xy}$ ($\rho_{xx}$) denotes
transverse (longitudinal) resistivity in the CuO$_2$ plane, crosses
over from a negative to a positive value near optimal doping,
implying that the charge character of the carriers changes from
being electron-like in the underdoped system to becoming hole-like
upon overdoping. For doping $x\le 0.14$ the $T-$dependence of
$\theta_H$ is seen in Fig. 4 to be monotonic and electron-like,
while for $x=0.19$ it is also monotonic but hole-like.  For
intermediate dopings, as illustrated by the data for $x$=0.15 and
0.16, it is non-monotonic and may change sign with increasing $T$,
from being hole-like at low $T$ to electron-like at high $T$.
Curiously, the Hall data at intermediate dopings can be fitted
reasonably via a weighted average of the experimental data at low
and high doping, some discrepancies notwithstanding. Specifically,
the solid line in Fig. 3 through the $x$=0.15 data points
corresponds to a superposition involving 90\% of the $x$=0.14 data
and 10\% of the $x$=0.19 data in the figure. The dashed line for the
$x$=0.16 dataset similarly corresponds to an admixture with 44\% of
the $x$=0.14 and 56\% of the $x$=0.19 dataset. These observations
leave little doubt that the intermediate doping regime involves two
carrier conduction. Note that the onset of two-carrier conduction
falls at essentially the same doping at which the nodal pockets
first appear (i.e. TTI) as determined from penetration depth
experiments. Very recently Li {\it et. al.}\cite{Li} have shown,
using a spin density wave model similar to ours, that a FS crossover
from a single electron-like FS to electron- and hole-like pockets is
consistent with an anomalous nonlinear magnetic field dependence of
the high field Hall resistivity.
\section{Discussion and Conclusion}
The importance of topological transitions in controlling the
properties of electron-doped cuprates has been considered by other
authors. Krotkov {\it et. al.}\cite{krotkov} have shown that above
what we have referred to as TTI the low energy spin dynamics is
dominated by gapless bosonic collective modes near the nodal point.
This causes the holes to develop a large mass, so that the resulting
{\it heavy fermion} behavior can alter the fermionic dynamics and
may lead to non-Fermi-liquid effects, including anomalous frequency
dependencies of the conductivity and Raman response. They also argue
that the strong reduction of $T_c$ in electron-doped cuprates
compared to the hole-doped case is a reflection of differences in
the FS topology. Another interesting consequence of the coexistence
of two carriers near optimal doping could be a BCS to Bose-Einstein
condensation (BEC) crossover. The generalized BEC
theory\cite{adhikari}, which includes the coexistence of
two-electron ($-2e$) and two-hole ($2e$) Cooper pairs associated
with separate bands, suggests the possibility that bosonic
electron$-$hole pairs could undergo Bose
condensation\cite{brinkman,adhikari}.

In conclusion, we have shown that our model of a uniformly doped
antiferromagnetic metal/superconductor is capable of describing the
doping evolution of a number of properties of the electron-doped
cuprates. Our model fundamentally involves the presence of two
distinct topological transitions of the FS, which are referred to as
TTI and TTII here. The existence of TTI (appearance of nodal hole
pockets) in the optimal doping region is indicated quite clearly by
the penetration depth and Hall effect experiments. TTII (merging of
the hole and electron pockets into a single large FS) lies close to
the solubility limit of the material and would be more difficult to
observe.

\begin{acknowledgments}
This work is supported by the U.S.D.O.E contracts DE-FG02-07ER46352
and DE-AC03-76SF00098 and benefited from the allocation of
supercomputer time at NERSC and Northeastern University's Advanced
Scientific Computation Center (ASCC).
\end{acknowledgments}


\begin{thebibliography}{99}
%
\bibitem{kusko} C. Kusko, {\it {et. al.,}} Phys. Rev. B. {\bf{66}},
140513(R) (2002).
%
\bibitem{tanmoy} Tanmoy Das, R. S. Markiewicz, and A. Bansil, Phys.
Rev. B. {\bf{74}}, 020506(R) (2006).
%
\bibitem{tanmoy2} Tanmoy Das, R. S. Markiewicz and A. Bansil, Phys.
Rev. Lett. {\bf{98}}, 197004 (2007).
%
\bibitem{foot6} We have extrapolated our order parameters
to the overdoped region and carried out self-consistent calculations
to estimate the QCP.
%
\bibitem{armitage} N. P. Armitage, {\it {et. al.,}} Phys. Rev. Lett.
{\bf{87}}, 147003 (2001).
%
\bibitem{markie} R.S. Markiewicz and A. Bansil, Phys. Rev. Lett. {\bf{96}},
107005 (2006).
%
\bibitem{anlage} S. M. Anlage, {\it {et. al.,}} Phys. Rev. B. {\bf{50}}, 523
(1994).
%
\bibitem{kashiwaya} S. Kashiwaya {\it {et. al.,}} Phys. Rev. B. {\bf{57}}, 8680
(1998).
%
\bibitem{skinta} J. A. Skinta, {\it {et. al.,}} Phys. Rev. Lett.,
{\bf{88}}, 207005 (2002).
%
\bibitem{biswas} A. Biswas, {\it {et. al.,}} Phys. Rev. Lett.
{\bf{88}}, 207004 (2002).
%
\bibitem{qazilbash} M. M. Qazilbash, {\it {et. al.,}} Phys. Rev. B.
{\bf{68}}, 024502 (2003).
%
\bibitem{chesca} B. Chesca, {\it {et. al.,}} Phys. Rev. B. {\bf{71}},
104504 (2005).
%
\bibitem{snezhko} A. Snezhko, {\it {et. al.,}} Phys. Rev. Lett.
{\bf{92}}, 157005 (2004).
%
\bibitem{ariando} Ariando, {\it {et. al.,}} Phys. Rev. Lett.
{\bf{94}}, 167001 (2005).
%
\bibitem{pcco} M.-S. Kim, {\it {et. al.,}} Phys. Rev. Lett.,
{\bf{91}}, 087001 (2003).
%
\bibitem{greene1} Y. Dagan, {\it {et. al.,}} Phys. Rev. Lett. {\bf{92}}, 167001
(2004).
%
\bibitem{greene2} Y. Dagan, and R. L. Greene, Phys. Rev. B {\bf{76}}, 024506
(2007).
%
\bibitem{krotkov} P. Krotkov and A.V. Chubukov, Phys. Rev. Lett.
{\bf{96}}, 107002 (2006); Phys. Rev. B {\bf{74}}, 014509 (2006).
%
\bibitem{brinkman} A. Brinkman, and H. Hilgenkamp, Physica C
{\bf{422}}, 71 (2005).
%
\bibitem{adhikari} S.K. Adhikari, {\it {et. al.,}} Physica C,
{\bf{453}}, 37-45 (2007).
%
\bibitem{mgreven} P. K. Mang, {\it {et. al.,}} Phys. Rev. Lett.
{\bf{93}}, 027002 (2004).
%
\bibitem{rosseinsky} M. J. Rosseinsky {\it {et. al.,}} Inorg. Chem.
{\bf{30}}, 2680 (1991).
%
\bibitem{tanmoy2gap} Tanmoy Das, R. S. Markiewicz, and A. Bansil,
arXiv:0711.0480.
%
\bibitem{prohammer} M. Prohammer and J. P. Carbotte, Phys. Rev. B.
{\bf{43}}, 5370 (1991).
%
\bibitem{foot5} Different groups find somewhat different dopings for
optimal $T_c$, e.g. M.-S. Kim {\it {et. al.,}}[\onlinecite{pcco}]
reports optimal doping near $x\approx 0.137$ for PCCO, where M. M.
Qazilbash {\it {et. al.,}} Phys. Rev. B. {\bf{72}}, 214510 (2005)
reports the optimal doping near $x\approx 0.145$ in both PCCO and
NCCO.
%
\bibitem{Li} Pengcheng Li, F. F. Balakirev, and R. L. Greene, Phys.
Rev. Lett. {\bf{99}}, 047003 (2007).
%
\end{thebibliography}
\end{document}